\documentclass[aps,twocolumn,superscriptaddress,showpacs]{revtex4}

\usepackage{epsfig}
\usepackage{color}
\usepackage{graphicx}
\usepackage{dcolumn}
\usepackage{bm}
\usepackage[normalem]{ulem}
\usepackage{textcomp}
\usepackage{soul}
\usepackage[dvipsnames]{xcolor}
\usepackage{amsmath}

\mathchardef\mhyphen="2D

\makeatother

\begin{document}

\title{
Effect of electronic correlations on the spectral and magnetic properties of ZrZn$_{2}$}

\author{S. L. Skornyakov}
\affiliation{M. N. Mikheev Institute of Metal Physics of Ural Branch of Russian Academy of Sciences, S. Kovalevskaya Street 18, 620990 Yekaterinburg, Russia}
\affiliation{Ural Federal University, 620002 Yekaterinburg, Russia}

\author{V. S. Protsenko}
\affiliation{M. N. Mikheev Institute of Metal Physics of Ural Branch of Russian Academy of Sciences, S. Kovalevskaya Street 18, 620990 Yekaterinburg, Russia}
\affiliation{Ural Federal University, 620002 Yekaterinburg, Russia}

\author{V. I. Anisimov}
\affiliation{M. N. Mikheev Institute of Metal Physics of Ural Branch of Russian Academy of Sciences, S. Kovalevskaya Street 18, 620990 Yekaterinburg, Russia}
\affiliation{Ural Federal University, 620002 Yekaterinburg, Russia}

\author{A. A. Katanin}
\affiliation{Moscow Institute of Physics and Technology, Institutsky lane 9, Dolgoprudny, 141700, Moscow region, Russia}
\affiliation{M. N. Mikheev Institute of Metal Physics of Ural Branch of Russian Academy of Sciences, S. Kovalevskaya Street 18, 620990 Yekaterinburg, Russia}

\date{\today}

\begin{abstract}
We present results of a theoretical study of a prototypical weak ferromagnet ZrZn$_2$. 
We use the density-functional theory (DFT)+dynamical mean-field theory (DMFT) method to 
study the electronic and local magnetic properties. The obtained DFT+DMFT electronic 
self-energies are Fermi-liquid like, indicating a small effective mass enhancement of 
the Zr $4d$ states $m^*/m\sim 1.1 - 1.3$ accompanied by partly formed local moments 
within the electronic states of $t_{2g}$ symmetry. The effect of electronic interaction 
is shown to be essential for determining the correct topology of some of the Fermi surface 
sheets. To study in detail the pressure dependence of the Curie temperature $T_{\rm C}$ 
and corresponding pressure-induced quantum phase transition, we consider an effective 
single-band model, constructed using the Zr $4d$ contribution to the total density of 
states. The model is studied within static and dynamic mean-field theory, as well as 
spin-fermion approach. We show that the spin-fermion approach yields the temperature 
dependence of susceptibility at ambient pressure and the pressure dependence $T_{\rm C}(p)$, 
including the first-order quantum phase transition at $p\approx 1.7$~GPa, comparable 
well with the experimental data.

\end{abstract}

 \maketitle

\section{INTRODUCTION}

ZrZn$_2$ is a well-known weak ferromagnet. Despite its magnetic properties have 
been studied since the 1950s~\cite{ZrZn2_1st}, their peculiarities are still actively 
debated. At ambient pressure this compound is ferromagnetic below the Curie 
temperature $T_{\rm C}\approx 30$~K. The ferromagnetism is, however, suppressed by 
pressure~\cite{ZrZn2-p} and disappears at $p\approx 1.65$~GPa \cite{ZrZn2-p1,ZrZn2-p2}. 
It was argued in Ref.~\cite{ZrZn2-QPT} that the quantum phase transition in ZrZn$_2$ 
under external pressure is in fact of the first order; at finite magnetic field the 
corresponding metamagnetic behavior is observed~\cite{ZrZn2-QPT,ZrZn2-Meta}. Near the 
quantum phase transition in zero magnetic field the exponent of the resistivity 
$\rho \propto T^\alpha$ changes \cite{ZrZn2-p2} from the value $\alpha=5/3$, which is 
characteristic for systems with ferromagnetic correlations \cite{Ogawa,Moriya}, to 
$\alpha=3/2$, characteristic for antiferromagnetic correlations. 

Density-functional theory (DFT) band structure calculations \cite{ZrZn2-BS1,ZrZn2-BS2} of 
ZrZn$_2$ revealed an extended van Hove singularity due to a flat part of dispersion near 
the $L$ point of the Brilloin zone, similar to that in nickel~\cite{Hausoel}. This flat part 
yields a peak of the density of states near the Fermi level~\cite{ZrZn2-BS2,ZrZn2-DOS}, 
characteristic to many ferromagnetic materials. This peak on one hand promotes ferromagnetism 
\cite{Wolfarth}, but on the other hand, it makes the Stoner theory even qualitatively 
inapplicable at finite temperatures, since competing channels of electron scattering 
become important in this situation, as it was studied actively in two-dimensional systems 
with van Hove singularities~\cite{2D-RG}. In general, Stoner theory predicts transition 
temperatures much larger than the corresponding experimental data and does not explain the 
linear temperature dependence of inverse susceptibility ~\cite{Moriya}, which points to the 
importance of correlation effects. The lowest-order paramagnon interaction was described 
within the Moriya theory~\cite{Moriya}. This theory was justified within the renormalization 
group method~\cite{Millis} and allowed to predict the universal behavior of weak itinerant 
magnets near quantum phase transitions. At the same time, for numerical predictions of 
non-universal properties it requires the knowledge of both the electronic dispersion 
and the (para)magnon spectrum. Moriya theory was also extended to include the effect of 
higher-order diagrams with respect to paramagnon interaction (expressed through the 
density of states for sufficiently large correlation lengths)~ \cite{HertzKlenin,Katanin,KataninIgoshev,KataninIgoshev1}. 

However, these spin-fluctuation approaches were not intensively applied to models 
with realistic densities of states; they also may be insufficient by the following 
reasons: (i) above mentioned interplay of different channels of electron scattering 
can yield a strong renormalization of particle-hole spin excitations by other channels 
of electron interaction; (ii) strong on-site correlations are not considered by 
these approaches. In particular, in the multi-orbital case the correlation-induced 
physics is more reach due to the Hund interaction which may trigger for example the 
Hund metal behavior~\cite{Hund,Hund1,Hund2}, characterized by (orbital-selective) 
formation of local moments. Although in weak magnets the effect of Hund exchange is 
expected to be less dramatic, than in strong magnetic materials, it should be also 
considered. 

Therefore, weak ferromagnets pose a challenge for theoretical studies since both, 
local and non-local correlations are expected to be essential to describe their 
properties. Accordingly, at least two aspects of weak itinerant magnetism should 
be investigated. First aspect is the role of strong local Coulomb interaction 
and peculiarities of realistic density of states for quasiparticle and magnetic 
properties. For that, state-of-the-art methods for calculation of electronic 
properties based on a combination of density functional theory and dynamical 
mean-field theory of correlated electrons 
(DFT+DMFT)~\cite{dftdmft_method,dftdmft_method1,dftdmft_method2} have shown to 
be a powerful theoretical tool for studying the physics of real materials~\cite{dftdmft_application}. 
Second aspect is the applicability of effective single-band models with realistic 
densities of states, and a possibility to use them to study magnetic properties 
of weak itinerant magnets. 

In this paper, we explore the effect of local Coulomb correlations on the electronic 
and magnetic properties of ZrZn$_2$ (space group $Fd\bar 3m$) within the DFT+DMFT 
method. We interpret the results of multi-orbital DFT+DMFT calcualtion within the 
effective one-band model constructed using the realistic density of states of 
ZrZn$_2$ and solved by DMFT. To understand details of the paramagnet to ferromagnet 
transition we furthermore study an effective one-band model within static mean field 
and spin-fermion model approaches.

The paper is organized as follows. In Sec.~\ref{dmft_method} we describe the technical 
details of our DFT+DMFT calculations. The corresponding results for spectral properties, 
local and non-uniform spin susceptibility and temperature dependence of the uniform spin 
susceptibility of ZrZn$_{2}$ are discussed in Sec.~\ref{results_5band}. The effective 
single-band model is considered in Sect.~ \ref{sec_1band_model} within DMFT 
(Sect. \ref{DMFT_1band_subsect}), mean-field approach (Sect.~ \ref{MF_subsect}) and 
spin-fermion model (Sect. \ref{sf_subsect}).  Finally, our results are summarized in 
Sec.~\ref{sec_conclusion}. In Appendix A we discuss behavior of the uniform particle-hole 
bubble at not too low temperatures, while in Appendix B we provide details on the 
equations of the spin-fermion model used to account the effect of spin fluctuations.

\section{DFT+DMFT Study}
\subsection{Method}
\label{dmft_method}
We first study the effect of local electronic correlations on the electronic structure 
and magnetic properties of ZrZn$_{2}$. For that we have employed the DFT+DMFT method 
implemented within the plane-wave pseudopotential approach with generalized gradient 
approximation (GGA) in DFT~\cite{GGA}. We use a basis set of Wannier functions constructed 
by means of the projection procedure~\cite{Wannier1,Wannier2} for an energy window 
spanning occupied Zn $3d$ and partially filled Zr $4d$ bands. The realistic DFT+DMFT 
many-body problem is solved by the continuous-time hybridization-expansion quantum 
Monte-Carlo method (CT-QMC, segment algorithm)~\cite{ctqmc}. In these calculations we 
neglect effects of spin-orbit coupling. The Coulomb interaction term has been treated 
in the density-density approximation with the average Hubbard interaction $U=2.5$~eV 
and Hund's exchange $J=0.3$~eV for the Zr $4d$ orbitals~\cite{U_in_Zr}. To account 
for Coulomb interaction energy already described by DFT we employ the fully-localized 
double-counting correction $V_{dc}=U(N_{d}-0.5)-0.5J(N_{d}-1)$ self-consistently 
evaluated from local occupations $N_{d}$. We have verified that using the 
around-mean-field form of the double counting term does not change essentially our 
results. Spectral functions and orbital-dependent band mass renormalizations were 
computed using the real-axis self-energy $\hat \Sigma(\nu)$ obtained from the Pad\'e 
analytical approximation of the DFT+DMFT imaginary-axis self-energy 
$\hat \Sigma(i\nu)$~\cite{Pade}. In order to determine the lattice parameter of cubic 
ZrZn$_2$ under pressure we compute the total energy as a function of volume using the 
GGA energy functional and shift the experimental lattice constant according to the 
third-order Birch-Murnaghan equation of state~\cite{Birch}. For simplicity, only 
hydrostatic contraction and expansion of the cubic unit cell are considered in these 
calculations.

Within DFT+DMFT we compute the uniform magnetic susceptibility as a derivative of the 
field-induced magnetization $M(T)$ with respect to the applied field $H$: 
\begin{equation}
\chi (T)=
\frac{\partial M(T)}{\partial H}=
\mu_{\rm B}^2\frac{\partial [n_{\uparrow }(T)-n_{\downarrow }(T)]}{\partial E_{h}},
\end{equation}
where $n_{\sigma}(T)$ is the occupation of the spin-$\sigma$ at a temperature $T$,  
$E_{h}=\mu_{\rm B} H$ is the bare splitting of electronic spectrum, and $\mu_{\rm B}$ 
is the Bohr magneton. In these calculations we check an absence of polarization in 
the zero field and the linear character of $M$ as a function of $E_{h}$.

\subsection{Results}
\label{results_5band}
\subsubsection{Electronic properties}
\begin{figure}[t]
\centering
\includegraphics[width=0.35\textwidth,clip=true,angle=-90]{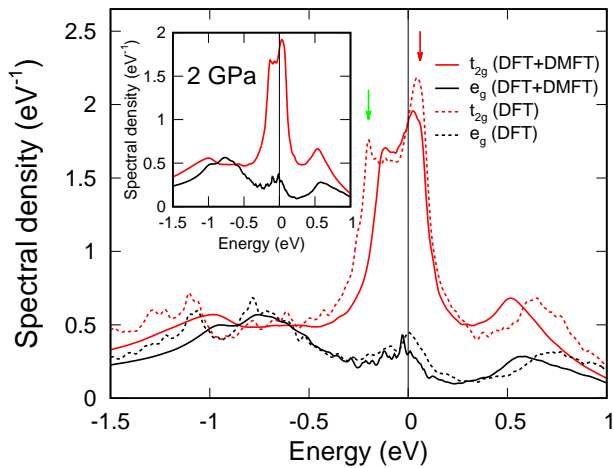}
\caption{
(Color online). Comparison of orbitally-resolved spectral functions of paramagnetic 
ZrZn$_{2}$ computed by DFT+DMFT ($T=464$~K) and DFT at ambient pressure. The inset 
shows DFT+DMFT results obtained at $p=2$~GPa. The Fermi energy is set to $0$~eV.
}
\label{Fig_1}
\end{figure}

As a starting point, we discuss the effect of electronic correlations on the electronic 
properties of ZrZn$_{2}$. The spectral functions of ZrZn$_{2}$ calculated by DFT and 
DFT+DMFT for the experimental crystal structure are presented in Fig.~\ref{Fig_1}. 
In agreement with previous theoretical investigations~\cite{ZrZn2-DOS,Kubler2004} our 
results (both DFT and DFT+DMFT) show that the spectral weight in the vicinity of the 
Fermi energy ($E_{\mathrm F}$) is due to the Zr $4d$ orbitals and mostly originates 
from the $t_{2g}$ states. These orbitals form a narrow band located in the interval 
$(-0.5,0.5)$~eV with sharp peaks below and above the $E_{\mathrm F}$ (marked by arrows 
in Fig.~\ref{Fig_1}). The $e_{g}$ spectral function also shows a broad peaked feature 
above the Fermi level, but its amplitude is almost five times smaller than that due to 
the $t_{2g}$ states. In addition, we observe that the Zn-$3d$ band is located well 
below $E_{\mathrm F}$ and is weakly hybridized with Zr-$4d$ states close to the Fermi 
level. 

\begin{figure}[b]
\centering
\includegraphics[width=0.35\textwidth,clip=true,angle=-90]{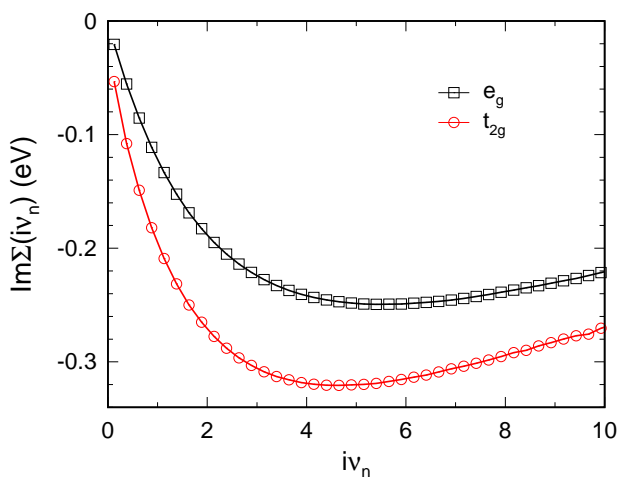}
\caption{
(Color online). Orbitally-resolved imaginary parts of the local self-energy 
of paramagnetic ZrZn$_{2}$ at ambient pressure on the Matsubara mesh obtained 
with DFT+DMFT ($T=464$~K).
}
\label{Fig_2}
\end{figure}

The Zr-$4d$ spectral functions computed by DFT+DMFT share a common shape with those 
obtained within DFT. Correlation effects only lead to a shift and renormalization of 
the quasiparticle bands near $E_{\mathrm F}$ and do not induce a significant transfer 
of the spectral weight. In particular, we observe that the sharp peak originating from 
the $t_{2g}$ states emerges at $\sim 0.025$~eV, almost twice closer to $E_{\mathrm F}$ 
compared to its position in DFT ($\sim 0.05$~eV). Such an effect of shifting the peak 
of the density of states towards the Fermi level is common for other correlated metallic 
systems. For example, it occurs in $\alpha$-~\cite{a-iron}, $\gamma$- \cite{OurGamma}, 
$\epsilon$-iron~\cite{e-iron}, FeAl~\cite{FeAl} and some two-dimensional systems (see, 
e.g., Refs.~\cite{Sr2RuO4,Skornyakov_PRB_2012,Skornyakov_PRB_2017}). The $e_{g}$ spectral 
function shows a similar transformation. However, the $e_{g}$-derived peak is pushed 
from above to below the Fermi energy. The obtained shape of the spectral functions is 
preserved in the whole pressure range $p<2$~GPa with all the features shifted to higher 
energies with increasing pressure. In particular, at $p=2$~GPa the peak of the $e_g$ 
states reaches the Fermi level.

To quantify electronic correlations in ZrZn$_{2}$ we analyze the local DFT+DMFT 
self-energy $\hat\Sigma(i\nu_{n})$ and the effective band mass enhancement 
$m^*/m=1-\partial {\rm Im}\hat\Sigma(i\nu)/\partial(i\nu))|_{i\nu\to 0}$. Here 
$i\nu_{n}$ is the fermionic Matsubara frequency and the derivative 
$\partial {\rm Im}\hat\Sigma(i\nu)/\partial(i\nu)$ is computed using Pad\'e 
extrapolation of $\hat\Sigma(i\nu)$ to $i\nu=0$. Our results for the frequency 
dependence of orbitally-resolved self-energies ${\rm Im}\hat\Sigma_{m}(i\nu_n)$ 
($m=t_{2g},e_{g}$) at an  electronic temperature $T=464$~K are presented in Fig.~\ref{Fig_2}. 
We observe that the $e_{g}$ self-energy exhibits a Fermi liquid-like behavior 
with insignificant damping of quasiparticles (${\rm Im}\hat\Sigma_{e_g}(0)\sim 0.002$~eV) 
and yields the mass enhancement $m^*/m=1.16$. The $t_{2g}$ states are less coherent 
and show a stronger renormalization with ${\rm Im}\hat\Sigma_{t_{2g}}(0)\sim 0.015$~eV 
and $m^*/m=1.34$. 

\begin{figure}[b]
\centering
\includegraphics[width=0.48\textwidth,clip=true,angle=0]{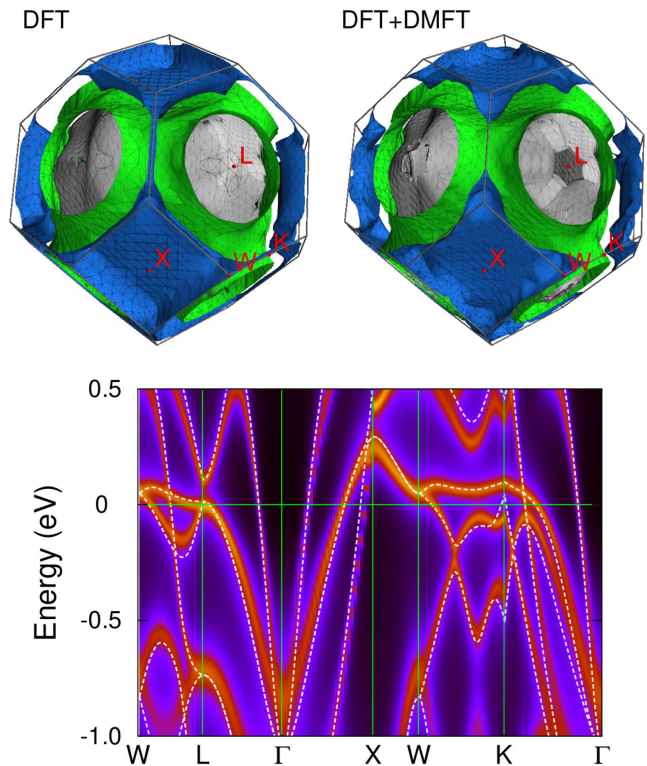}
\caption{
Upper panel: Fermi surface of ZrZn$_{2}$ at ambient pressure as obtained within 
DFT (left) and that computed by DFT+DMFT at $T=464$~K (right). Lower panel: Band 
structure of paramagnetic ZrZn$_{2}$ computed by DFT+DMFT at $T=464$~K (contours) 
and DFT (dashed lines). 
}
\label{Fig_3}
\end{figure}

We proceed further with investigation of the effect of electronic correlations on the 
Fermi surface (FS) and band structure of ZrZn$_{2}$. To compute the FS within DFT+DMFT 
we locate poles of the analytically continued lattice Green function $G_{\bf k}(\nu=0)$, 
diagonalized in the orbital space. In Fig.~\ref{Fig_3} (upper panel) we compare our 
results for the FS calculated by DFT and DFT+DMFT at ambient pressure. Both 
DFT and DFT+DMFT 
approaches yield the FS that shows four sheets centered at the $\Gamma$ point. Specifically, 
the FS consists of the innermost spherical sheet (not seen in Fig.~\ref{Fig_3}) 
surrounded by three other sheets with more complex shape. We observe that upon inclusion 
of electronic correlations topology of the inner sheet (Fig.~\ref{Fig_3}, grey) changes 
near the $L$ point and the outer sheet (Fig.~\ref{Fig_3}, blue) splits along 
the $W-K$ direction of the Brillouin zone, such that the DFT+DMFT Fermi surfaces are 
in agreement with experimental data \cite{FS}. Analysis of the band structure 
(Fig.~\ref{Fig_3}, lower panel) shows that this transformation occurs due to a 
correlation-induced shift of the bands at the $L$ and $K$ points of the Brilloiun zone, 
from above to below the Fermi level. Therefore, correlations are crucially important 
to obtain the correct topology of the Fermi surface of ZrZn$_2$ (see also discussion 
in Ref.~\cite{FS}). 
\subsubsection{Local spin susceptibility\label{SectSuscLoc}}
\begin{figure}[t]
\centering
\includegraphics[width=0.7\textwidth,clip=true,angle=-90]{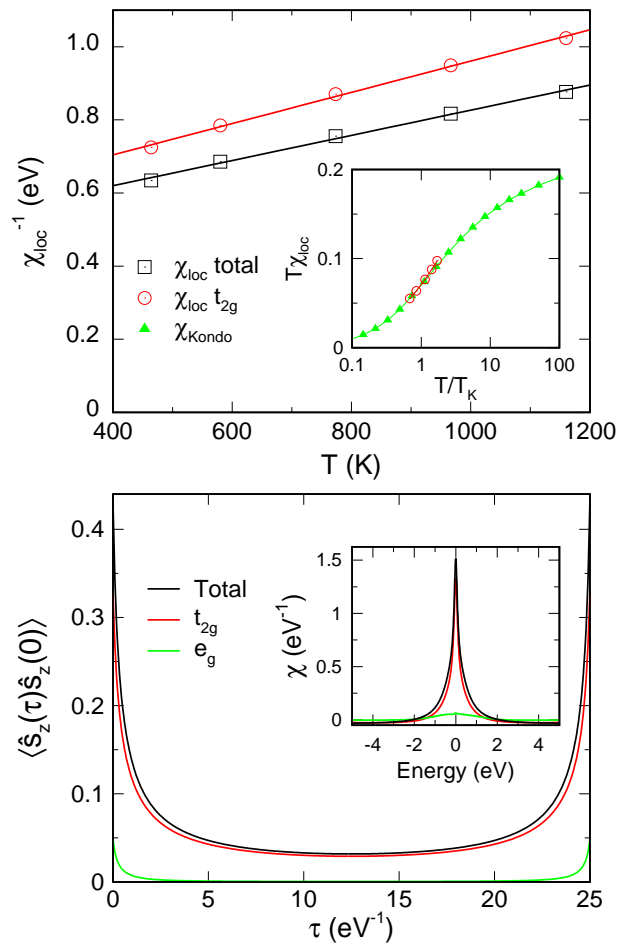}
\caption{
(Color online). Upper panel: Temperature dependence of the inverse local 
spin susceptibility $\chi_{\mathrm{loc}}^{-1}$ of paramagnetic ZrZn$_{2}$ 
and the inverse of the corresponding $t_{2g}$ contribution, computed by DFT+DMFT. The inset 
shows a fit of the $t_{2g}$ contribution to local susceptibility (circles) 
by the universal dependence $T\chi_{\rm loc}(T/T_{\rm K})$ for the Kondo 
model~\cite{Kondo} (triangles), allowing to obtain the Kondo temperature 
$T_{\rm K}$. Lower panel: Local spin correlation function 
$\chi(\tau)=\left<\hat s_i^{z}(\tau)\hat s_i^{z}(0)\right>$ of paramagnetic 
ZrZn$_{2}$ as computed by DFT+DMFT at $T=464$~K. Orbitally-resolved Fourier 
transform of $\chi(\tau)$ as a function of the real energy is shown in the 
inset.
}
\label{Fig_4}
\end{figure}

We consider further the local spin susceptibility 
$\chi(\tau)=\langle \hat s_{i}^z(\tau)\hat s_i^{z}(0)\rangle$ (where $\hat s_i^{z}(\tau)$ 
is the instantaneous spin at the site $i$ at imaginary time $\tau$ and $\langle \ldots\rangle$ 
denotes the thermal average computed by CT-QMC) and its Fourier transform  
$\chi_{\rm loc}(i\omega_n)=\int_{0}^{1/k_{\rm B}T}\chi(\tau)\exp(i\omega_n \tau)d\tau$, 
$\omega_n$ being bosonic Matsubara frequencies. Our results for the temperature 
dependence of the inverse static local spin susceptibility 
$\chi_{\rm loc}^{-1}=(\chi_{\rm loc}(0))^{-1}$ are shown in Fig.~\ref{Fig_4} 
(upper panel). Calculated $\chi_{\mathrm{loc}}^{-1}$ and the corresponding 
inverse partial contributions of $t_{2g}$ and $e_g$ states to a good approximation 
are linear functions of temperature. The contribution of $e_g$ states to the 
susceptibility (not shown) is much smaller than that of the $t_{2g}$ states. 
From the fit of the $t_{2g}$ contribution by the universal temperature dependence 
$T\chi_{\rm loc}(T/T_{\rm K})$ of the impurity susceptibility of the Kondo model~\cite{Kondo}, 
we find a rather large Kondo temperature $T_{\rm K}\approx 680$~K, which shows that 
local moments are fully screened at low temperatures. The presence of a flat part of 
time dependence of $\chi(\tau)$ (see lower panel of Fig.~\ref{Fig_4}) is indicative 
of short-lived local moments in the $t_{2g}$ band. Their inverse lifetime is 
characterized by the half width of the peak of the analytical continuation of 
$\chi(\omega)$ at the half height~\cite{OurGamma} and it is equal to $0.16$~eV. This value corresponds 
to the lifetime $25$~fs, which is approximately twice longer than that discussed 
previously for the iron pnictides~\cite{Pnictides} and $\epsilon$-iron~\cite{e-iron}, 
but much shorter than the lifetime of local moments in such strong magnet as 
$\alpha$-iron \cite{a-iron}. At the same time, almost no local moments are formed 
in the $e_g$ states. These results are in line with orbital-selective behavior of 
the self-energy and imply a different degree of electronic coherence for different 
orbitals of the Zr $4d$ shell, indicative for orbital-selective local moments.

\vspace{2cm}

\vspace{-0.25cm}
\subsubsection{Momentum dependence of the non-uniform magnetic susceptibility}

\begin{figure}[t]
\centering
\includegraphics[width=0.33\textwidth,clip=true,angle=-90]{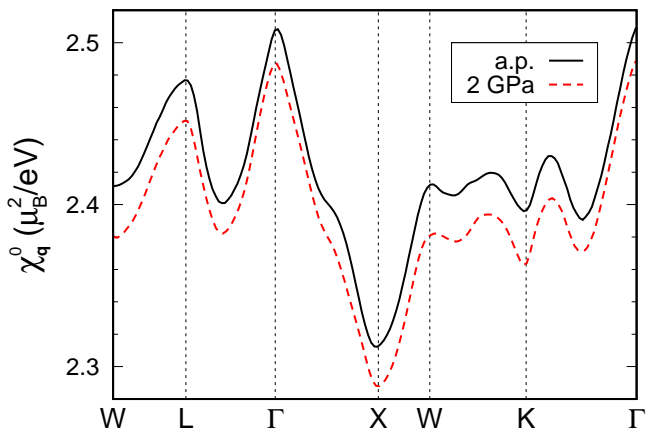}
\caption{(Color online). Momentum dependence of the particle-hole bubble 
$\chi_\textbf{q}^{0}$ in ZrZn$_2$ computed by DFT+DMFT at ambient pressure 
(a.p.) and $2$~GPa.
}
\label{Fig_5}
\end{figure}

To study preferable types of magnetic correlations in ZrZn$_2$, in Fig.~\ref{Fig_5} 
we show the momentum dependence of the nonuniform magnetic susceptibility, 
represented by the particle-hole bubble,
\begin{equation}
\chi_\textbf{q}^{0} = -(2\mu_{\rm B}^2/\beta) \sum_{\textbf{k},\nu_n} \textrm{Tr}[\hat{G}_\textbf{k} (i\nu_n) \hat{G}_\textbf{k+q}(i\nu_n)],\label{Eq:Bubble}
\end{equation}
where $\hat{G}_\textbf{k}(i\nu_n)$ is the one-particle DFT+DMFT Green function 
on Zr atom, which is a matrix in the space of $d$-orbitals, $\beta=1/k_{\mathrm B}T$. 
One can see that apart from the peak at ${\bf q}_\Gamma=0$, which shows a tendency 
to ferromagnetic order, there is a strong peak of $\chi_\textbf{q}^{0}$ at 
${\bf q}_L={(\pi,\pi,\pi)}$, showing the presence of spin density wave correlations. 
Detail analysis of contribution of different momenta shows that this peak originates 
from the nesting between wide necks of the FS sheets close to the $W-K$ edge of the 
Brillouin zone (Fig.~\ref{Fig_3}, green) and corners of the pillow-like FS centered 
at the $X$ points (Fig.~\ref{Fig_3}, blue).

The tendency towards spin density wave may explain the experimentally observed $T^{3/2}$ 
dependence of the resistivity in the paramagnetic state under pressure~\cite{ZrZn2-p2}.

\subsubsection{Uniform magnetic susceptibility}
\begin{figure}[t]
\centering
\includegraphics[width=0.35\textwidth,clip=true,angle=-90]{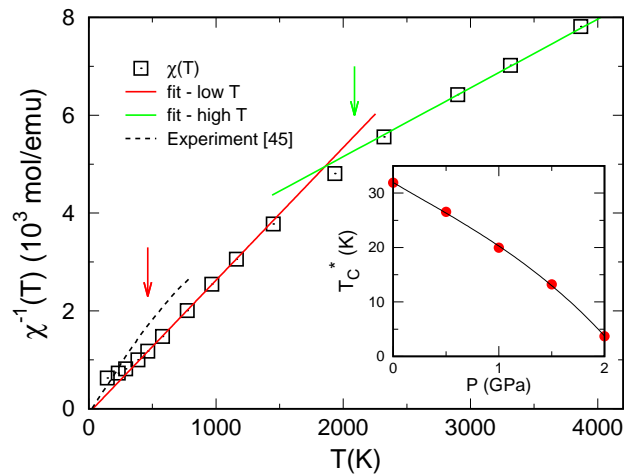}
\caption{
(Color online). Temperature dependence of the inverse uniform spin susceptibility 
$\chi^{-1}(T)$ of paramagnetic ZrZn$_{2}$ computed by DFT+DMFT at ambient pressure 
(squares) in comparison with experiment (broken line)~\cite{Shimizu_1981}. The straight 
solid lines represent least-square fits to $\chi^{-1}(T)$ in the temperature range 
$[400,1200]$~K and $[2000,4000]$~K. Temperatures corresponding to excitation energies 
of the peaks in the density of states are indicated by arrows. The inset shows the 
pressure dependence of $T_{\rm C}^{*}$ as calculated from the low-temperature fit 
of $\chi^{-1}(T)$.
}
\label{Fig_6}
\end{figure}

To complement the analysis of magnetic properties of ZrZn$_2$ we consider the 
temperature dependence of inverse uniform magnetic susceptibility $\chi^{-1}(T)$. 
In Fig.~\ref{Fig_6} we compare our DFT+DMFT results for $\chi^{-1}(T)$ to experimental 
data of Shimizu {\it et al.}~\cite{Shimizu_1981}. To highlight the regions of Curie-Weiss 
behavior, we show least square linear fits to the computed points. One can see that the 
Curie-Weiss law is fulfilled at $T>T_{\rm CW}\approx 400$~K; the slope of inverse susceptibility 
changes at the temperature $T'_{\rm CW}\approx 2000$~K. The temperatures $T_{\rm CW}$ 
and $T'_{\rm CW}$ approximately correspond to the energy of the two peaks of the DFT 
$t_{2g}$ density of states, closest to the Fermi level ($\varepsilon _1\approx-0.15$~eV 
and $\varepsilon _2\approx 0.05$~eV, respectively), cf. Fig.~\ref{Fig_1}. The reason for 
the change of magnetic susceptibility at $T=T_{\rm CW}$ and $T'_{\rm CW}$ is that the 
electronic correlations are strongly enhanced by peaks of the density of states.

At $T>T'_{\rm CW}$ we find a magnetic moment $\mu=2.38\mu_{\rm B}$. This value 
corresponds to weakly interacting electrons ($\mu=2.39\mu_{\rm B}$ is obtained from 
the lowest order bubble contribution (\ref{Eq:Bubble}) for Zr states, relevant in the 
considered temperature range, see Appendix A). At the same time, at 
$T_{\rm CW}<T<T'_{\rm CW}$ we find a reduced magnetic moment $\mu=1.72\mu_{\rm B}$.
Via the relation 
\begin{equation}
\mu^2=(g\mu_{\rm B})^2p(p+1),\label{Effspin}
\end{equation} 
where $g=2$, this value corresponds 
to an effective spin $p=0.49$. This value can be interpreted as corresponding to a 
correlated electronic state with partial localization of one hole in the $t_{2g}$ 
band (filling $n_{t_{2g}}=4.5$), which agrees with the discussion in Sect.~\ref{SectSuscLoc}.

The slope of $\chi^{-1}(T)$ in the range $T_{\rm CW}<T<T'_{\rm CW}$, and, therefore, 
the respective magnetic moment, is in agreement with experimental data, although with 
some shift of the inverse susceptibility, which origin is to be clarified by future 
investigations. Orbitally-resolved contributions $\chi_{\alpha}$ ($\alpha=$ $t_{2g}$, 
$e_{g}$) to the total susceptibility (not shown) indicate that the temperature evolution 
of $\chi(T)$ in a wide temperature range mostly originates from the $t_{2g}$ orbitals.
The contribution of the $e_{g}$ states is an order of magnitude smaller and becomes 
notable only in the low-temperature region. 

At lower temperatures $T<T_{\rm CW}\sim T_{\rm K}$ theoretical $\chi^{-1}(T)$ dependence 
shows an upturn, which is likely related to the Kondo screening of the local moment 
formed by $t_{2g}$ states. Such upturn is found in DFT+DMFT studies of many Hund metals 
with van Hove singularity, slightly shifted off the Fermi level \cite{OurGamma,LaFeAsO,Skornyakov_PRB_2012,e-iron,FeAl}, although it is not observed in 
the experimental temperature dependence of the susceptibility of ZrZn$_2$. The obtained 
upturn of $\chi^{-1}(T)$ at low temperatures physically implies a tendency towards a 
non-ferromagnetic ground state. Our results demonstrate that this tendency, which 
originates from the Kondo screening of local moment, formed by the $t_{2g}$ states, is 
overestimated in the DMFT approach for ZrZn$_2$. As a result, DFT+DMFT is applicable 
to the considered system only at not too low temperatures.

To proceed further with the results of DFT+DMFT approach, we extract a characteristic 
temperature  $T_{\rm C}^*$ of the onset of strong ferromagnetic correlations 
(which is approximately identified with the Curie temperature $T_{\rm C}$) by 
extrapolating the dependence of susceptibility at high temperatures according to 
the Curie-Weiss law, $\chi^{-1}\propto T-{ T_{\rm C}^*}$. The extrapolation of the 
fit to $\chi^{-1}=0$ at ambient pressure gives the temperature ${ T_{\rm C}^*}\sim 32$~K, 
close to the value reported in Ref.~\cite{Shimizu_1981}. 

Upon increasing pressure we observe a growth of $\chi^{-1}(T)$ in the temperature 
range $T<1200$~K. The effect of pressure is accompanied by a monotonic reduction of 
${T_{\rm C}^*}$ (Fig.~\ref{Fig_6}, inset). In particular, at $p=1$~GPa we obtain a drop 
of ${ T_{\rm C}^*}$ to $\sim 20$~K which further develops to $\sim 4$~K at $p=2$~GPa. 
This behavior is in overall qualitative agreement with experimental tendency of 
decreasing the Curie temperature under compression of the 
lattice~\cite{ZrZn2-p1,ZrZn2-p2,ZrZn2-QPT,ZrZn2-Meta}.

Reconciling the results of DFT+DMFT approach with the experimental data at low 
temperatures requires consideration of  non-local correlations. The treatment of 
non-local spin correlations within the spin-fermion model is presented below, in 
Sect. \ref{sf_subsect}. In Sect. IV we also discuss possible reasons of inapplicability 
of DMFT in the low-temperature region.

\section{Effective single-band model}
\label{sec_1band_model}
\subsection{The model and the dynamical mean-field theory\label{DMFT_1band_subsect}}
\begin{figure}[t]
\centering
\includegraphics[width=0.75\textwidth,clip=true,angle=-90]{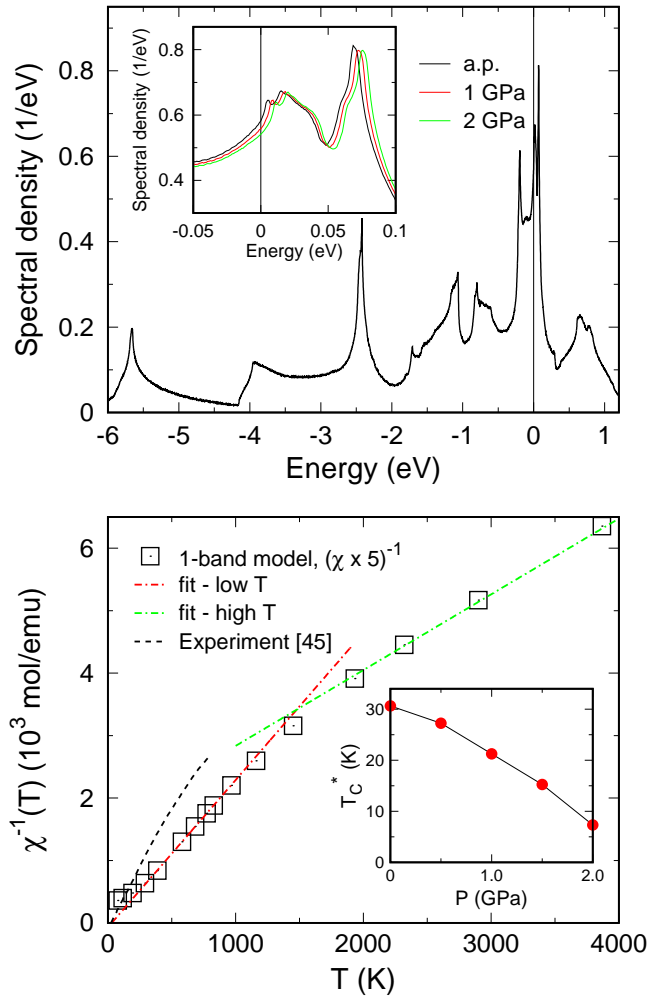}
\caption{
(Color online). Top: non-interacting density of states of a one-band model at 
ambient pressure (a.p.) and its evolution close to the Fermi level ($0$~eV) 
at $p=1$~GPa and $p=2$~GPa (inset). Bottom: Temperature dependence of the 
inverse uniform susceptibility $\chi^{-1}(T)$ of the one-band model as computed 
by DMFT at ambient pressure using the Coulomb parameter $U=4.5$~eV (squares). 
The dot-dashed lines show linear fits, the dashed line corresponds to the 
experimental data of Ref. \cite{Shimizu_1981}. The inset shows the pressure 
dependence of $T_{\rm C}^*$ extracted from a linear fit to the low-temperature 
part of $\chi^{-1}(T)$.
}
\label{Fig_7}
\end{figure}

To study the phase transition to ferromagnetic phase in more detail, we consider an 
effective single-band model. To construct this model we use a cut of the realistic 
total density of states of ZrZn$_2$ computed by DFT in an energy window chosen to 
span the full width of Zr-$4d$ band. The normalized density of states $\rho(\epsilon)$ 
(Fig. \ref{Fig_7}, upper panel) defines a single-band model which is solved by dynamical 
and static mean-field theory, as well as investigated within the spin-fermion model 
approach. In these calculations the occupation at each pressure is fixed at its value 
obtained by using the Fermi level position in DFT calculation (e.g. at $p=0$~GPa after 
normalization this procedure yields a total occupation $n=1.56$, which is close to 
the averaged occupation of Zr-$d$ Wannier states).

In the dynamical mean-field theory we consider a standard single-impurity Anderson 
model (AIM), supplemented by the self-consistency condition 
$\int \rho(\epsilon)/(i\nu_n-\epsilon-\Sigma(i\nu_n))=g(i\nu_n)$ where  
$g(i\nu_n)$ is the local Green's function and $\Sigma(i\nu_n)$ is the local self-energy, 
obtained within the AIM. We note that previously a similar approach was used for study 
a ferromagnetic instability in systems with model densities of states~\cite{Held}. 
Here, we use the ab initio density of states $\rho(\epsilon)$, constructed as discussed 
above. To compare our DMFT results for the single-band model with those of Sect.~\ref{results_5band} 
for a realistic multi-band model, we use in the DMFT approach to the single band model 
the local Coulomb interaction $U=4.5$~eV chosen such that the value of characteristic 
temperature $T_{\rm C}^*$ at ambient pressure is close to that obtained for the multi-band 
model.

The corresponding DMFT results for the inverse spin susceptibility, multiplied by a 
number of relevant orbitals of the real compound, i.e. $(5\chi(T))^{-1}$, are shown 
in  Fig.~\ref{Fig_7}, lower panel. These results are quite similar to those of the 
multi-orbital model. In particular, in the low-temperature region the computed 
inverse susceptibility shows an upturn in the low-temperature region. We note that 
similar upturn was obtained also in earlier studies of the single-band Hubbard model 
with the peak of the density of states shifted off the Fermi level~\cite{Skornyakov_PRB_2012}. 
By analogy to the multi-orbital case we extract the characteristic temperature 
$T_{\rm C}^{*}$ of the onset of strong ferromagnetic correlations from a linear fit 
of the low-temperature part of $\chi^{-1}(T)$. At ambient pressure our calculations 
yield a $T_{\rm C}^{*}\sim 30$~K which drops to $\sim 7$~K at $p=2$~GPa in qualitative 
agreement with the realistic DFT+DMFT calculation (Fig.~\ref{Fig_7}, lower panel, inset). 
Studying the paramagnetic to ferromagnetic phase transition requires an account for 
non-local correlations supplemented by a comparison of energies of para- and ferromagnetic 
states. We note that importance of non-local correlations for stability of ferromagnetism 
was emphasized earlier in the DMFT study of a single-band model in Ref.~\cite{Held}. 
As we show in the following subsections the transition to the ferromagnetic state can 
be successfully described by employing a simplified spin-fermion model.

\subsection{Mean-field approximation\label{MF_subsect}}
Before analyzing magnetic properties of the single-band model within the 
spin-fermion approach, we consider the static mean-field approximation, 
described by the action
\begin{multline}
\mathcal{S}_{{\rm MF}}=\sum\limits_{k,\sigma} c_{k,\sigma}^{\dagger}\left(-i\nu_{n}+\epsilon_{\textbf{k}
}-\mu\right)c_{k,\sigma}+U(n_{\uparrow}n_{\downarrow})_{\rm MF},
\label{action_mf}
\end{multline}
where $c_{k,\sigma}^{\dagger},c_{k,\sigma}$ are Grassmann variables, $\nu_n$ are 
fermionic Matsubara frequencies,
\begin{eqnarray}
(n_{\uparrow}n_{\downarrow})_{\rm MF}&=&\langle n_{\uparrow} \rangle n_{\downarrow}+\langle n_{\downarrow} \rangle n_{\uparrow}-\langle n_{\uparrow} \rangle \langle n_{\downarrow}\rangle\notag \\
&=&n\langle n \rangle /{2}-2\langle s_{z}\rangle s_{z}-\langle n \rangle^2/4+\langle s_z \rangle^2,
\end{eqnarray}
$n=n_\uparrow+n_\downarrow$, $s_z=(n_\uparrow-n_\downarrow)/2$, 
$n_\sigma=\sum_k c_{k,\sigma}^{\dagger} c_{k,\sigma}$, $k=(\textbf{k},\nu_{n})$ is 
the momentum-frequency 4-vector, $\epsilon_{\textbf{k}}$ is an electronic dispersion 
of the one-band model, and $\mu$ is the chemical potential. This leads to  well-known 
self-consistent equations for $\langle n\rangle$ and $\langle s_{z}\rangle$,
\begin{subequations}
\begin{eqnarray}
\langle n \rangle &=&\sum_{\sigma}{\int{f\left(\tilde{\epsilon}+\sigma U\langle s_{z}\rangle\right)\rho(\epsilon)d\epsilon}},\\
\langle s_{z} \rangle&=&-\frac{1}{2}\sum_{\sigma}{\sigma\int{f\left(\tilde{\epsilon}+\sigma U\langle s_{z}\rangle\right)\rho(\epsilon)d\epsilon}},
\label{Sz}
\end{eqnarray}
\label{MF_eqns}
\end{subequations}
where $\tilde{\epsilon}=\epsilon-\mu+{U\langle n \rangle}/{2}$, {$\sigma=\pm 1$,} 
$f(\epsilon)=(1+\exp{(\beta \epsilon)})^{-1}$ is the Fermi function,
and $\rho\left(\epsilon\right)$ is the density of states per one spin projection.

We solve the equations (\ref{MF_eqns}) for $\mu$ and $\langle s_z \rangle$ for a given 
electron concentration per orbital $\langle n \rangle$, obtained in DFT calculations. 
This yields paramagnetic ($\mu_{\rm PM}$, $\langle s_z \rangle=0$) and ferromagnetic 
($\mu_{\rm FM}$, $m_{\rm FM}=\langle s_z \rangle\neq 0$) solutions. To find energetically 
preferable solution, we compare the values of thermodynamic potential
\begin{eqnarray}
    \Omega(\mu;\langle n\rangle,\langle s_z\rangle)&=&U\left(\langle s_{z}\rangle^{2}-\langle n\rangle^{2}/4\right)\label{Omega_MF}\\
    -\beta^{-1}\sum_{\sigma}\int &\ln&{\left[1+e^{-\beta\left(\tilde{\epsilon}+\sigma U\langle s_{z}\rangle\right)}\right]}\rho(\epsilon)d\epsilon. \notag
\end{eqnarray}
If $\Omega(\mu_{\rm PM};\langle n\rangle,0)<\Omega(\mu_{\rm PM};n_{\rm PM},\langle s_z\rangle_{\rm PM})$, 
where $n_{\rm PM}$ and $\langle s_z\rangle_{\rm PM}\neq 0$ fulfill the mean-field equations for 
$\mu=\mu_{\rm PM}$, then the paramagnetic solution is considered energetically preferable, while 
in case $\Omega(\mu_{\rm FM};\langle n\rangle,m_{\rm FM})<\Omega(\mu_{\rm FM};n_{\rm FM},0)$, 
where $n_{\rm FM}$ and $\langle s_z\rangle=0$ is the solution of MF equations for 
$\mu=\mu_{\rm FM}$, the ferromagnetic solution dominates.

Solution of the equations (\ref{MF_eqns}) and (\ref{Omega_MF}) in the ground state for 
moderate Coulomb interaction shows that the ferromagnetic ground state is energetically 
preferable for positions of the peak of the density of states close to the Fermi level 
(sufficiently low pressures), while paramagnetic state is energetically preferable for 
peak positions far from the Fermi level. 

We note that the effective interaction $U$ in the static mean-field theory is typically 
smaller than the bare Coulomb interaction of the one-band model, since it accounts for 
the screening processes, cf. Refs.~\cite{Kanamori,KataninYamase}. We choose below the 
effective Coulomb interaction $U=1.8$~eV from the condition that the first-order quantum 
phase transition occursat the pressure $p_c$ close to the experimental value $p_c=1.65$~GPa.

\begin{figure}[t]
\centering
\vspace{0.2cm}
\includegraphics[width=0.9\linewidth]{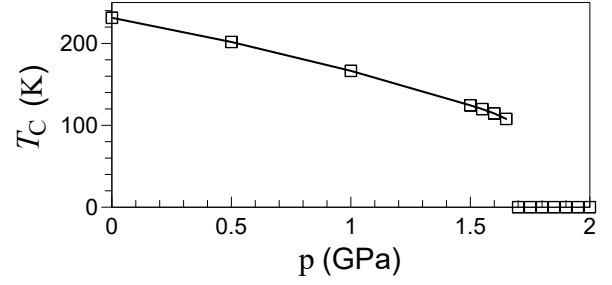}
\caption{
Pressure dependence of the Curie temperature $T_{\rm C}(p)$ in the mean-field analysis 
of the single-band model for $U=1.8$~eV. 
}
\label{TC_MF}
\end{figure}

The calculated dependence of Curie temperature $T_{\rm C}(p)$ in the mean-field theory 
is presented in Fig. \ref{TC_MF}. The obtained pressure dependence of $T_{\rm C}$ is in 
qualitative agreement  with the experimental data~\cite{ZrZn2-QPT,ZrZn2-Meta}. However, 
the obtained Curie temperatures overestimate experimental values by an order of magnitude, 
which is usual for mean-field theory of weak magnetic systems~\cite{Moriya}. In the narrow 
region of pressures near quantum phase transition, where none of the above criteria of 
preference of paramagnetic or ferromagnetic solution is fulfilled, the phase separation 
between para- and ferromagnetic phases occurs. We do not study this region in details 
since it is rather narrow for ZrZn$_2$.

\subsection{Spin-fermion model\label{sf_subsect}} 
To improve the results of the mean-field approximation, let us take into account the 
effect of spin fluctuations in the spin-fermion model. To this end, we consider the 
partition function 
\begin{equation}
Z=\int{\mathcal{D}[c,c^{\dagger}]\int{d^{3}{\tilde S}}\exp{\left(-\beta \mathcal{S}_{{\rm eff}}\right)}}
\label{PF}
\end{equation}
with the effective action
\begin{equation}
\mathcal{S}_{{\rm eff}}=\mathcal{S}_{\rm MF}+2U\textbf{s}\tilde{\textbf{S}}+D\tilde{\textbf{S}}^{2},
\label{action}
\end{equation}
containing the fluctuating field $\tilde{\textbf{S}}$;  $\textbf{s}=({1}/{2})\sum_{k,\sigma,\sigma^{'}}{c_{k,\sigma}^{\dagger}\mbox {\boldmath $\sigma $}_{\sigma\sigma^{'}}c_{k,\sigma^{'}}}$
corresponds to the spin of itinerant degrees of freedom, 
$\mbox {\boldmath $\sigma $}=(\sigma_{x},\sigma_{y},\sigma_{z})$ is the vector of Pauli 
matrices, $D$ is the strength of spin fluctuations. For simplicity we consider here only 
the effect of spin fluctuations of static uniform (${\bf q}=0$, $\omega_n=0$) spin mode, 
which can be justified at sufficiently large correlation 
lengths~\cite{HertzKlenin,Schmalian,Katanin,KataninIgoshev,KataninIgoshev1}. We also note 
that at finite temperatures the classical fluctuations in dimension $d=3$ are relevant from 
renormalization group point of view, in contrast to quantum fluctuations, having dimension 
$d+z$, where $z=3$ is the dynamic critical exponent \cite{Millis}. To be consistent with 
the mean-field action (\ref{action_mf}), we keep the spin-fermion interaction equal to 
$2U$, cf. Ref.~ \cite{KataninIgoshev}. This yields the following equations for the electronic 
density and the magnetization in the absence of magnetic field (see Appendix B)

\begin{eqnarray}
\langle n \rangle &=&
A\int{d^{3}}\tilde{S}\exp\left(-\beta D\tilde{\textbf{S}}^{2}\right)\label{SF_EQn}\\&\times&\sum_{\sigma}{\int{f\left(\tilde{\epsilon}+\sigma\gamma\right)\rho(\epsilon)d\epsilon}},\notag\\
\langle s_{z} \rangle&=&-\frac{AU}{2}\int{d^{3}}\tilde{S}\exp\left(-\beta D\tilde{\textbf{S}}^{2}\right)\dfrac{\langle s_{z}\rangle-\tilde{S}_z}{\gamma}\label{SF_EQm}\\&\times&\sum_{\sigma}{\sigma\int{f\left(\tilde{\epsilon}+\sigma\gamma\right)\rho(\epsilon)d\epsilon}},\notag
\end{eqnarray}
where 
$\gamma=U(\langle s_{z}\rangle^{2}-2\langle s_{z}\rangle\tilde{S}_z+\tilde{S}^{2})^{1/2}$, $A=({\beta D}/{\pi})^{3/2}$.

\begin{figure}[t]
\centering
\includegraphics[width=0.9\linewidth]{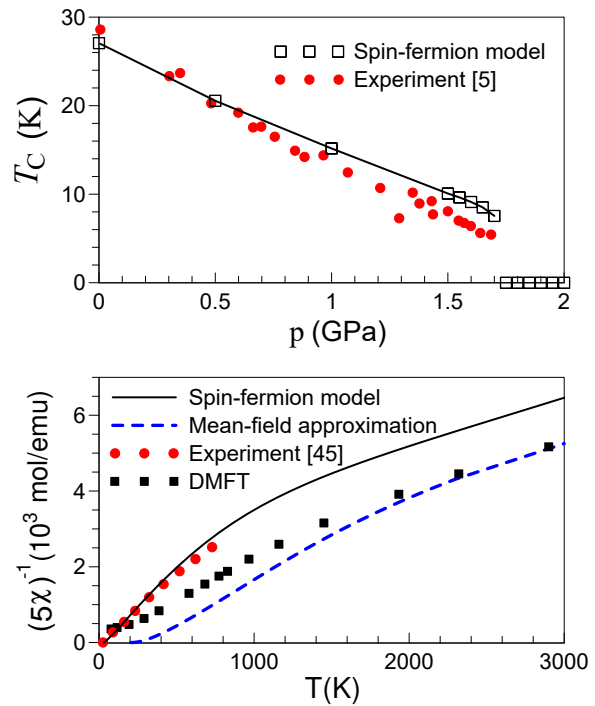}
\caption{
(Color online). Magnetic properties of ZrZn$_2$ obtained within spin-fermion 
model (solid lines) in comparison to the experimental data (circles). 
Upper panel: Pressure dependence of the Curie temperature $T_{\rm C}(p)$. 
The experimental data are taken from Ref. \cite{ZrZn2-QPT}. 
Lower panel: Temperature dependence of inverse susceptibility, rescaled 
with the number of orbitals, $(5\chi(T))^{-1}$ at ambient pressure, compared 
to the respective results of the mean-field approximation (dashed line), and 
DMFT for single-band model (squares, cf. Fig. \ref{Fig_7}). The experimental 
data are taken from Ref.~\cite{Shimizu_1981} (without rescaling).
}
\label{TC_SF}
\end{figure}

For numerical solution of equations {(\ref{SF_EQn}) and (\ref{SF_EQm})} we keep the value 
$U=1.8$~eV which was chosen in Sect.~\ref{MF_subsect} for the mean-field analysis and choose 
temperature-independent $D^{-1}=0.15$~eV$^{-1}$, which allows us to achieve the Curie 
temperature at ambient pressure, which is approximately equal to the experimental value. 
As in the mean-field approach we solve first the system of equations {(\ref{SF_EQn}) and 
(\ref{SF_EQm})} for given concentration of electrons $\langle n\rangle$, obtained in DFT, 
and then compare thermodynamic potentials $\Omega$ of para- and ferromagnetic solutions with 
the same chemical potential (see Appendix B for the explicit expression of $\Omega$ for the 
spin-fermion model).

The obtained dependence  of Curie temperature $T_{\rm C}(p)$ (see Fig. \ref{TC_SF}) 
is almost linear, and demonstrates a quantitative agreement with the experimental 
data~\cite{ZrZn2-QPT,ZrZn2-Meta}. The position of the first-order quantum phase 
transition $p_c$ is only weakly changed in comparison to the mean-field theory, and 
therefore remains also close to the experimental value. Therefore, we find that static 
almost uniform classical spin fluctuations are sufficient to successfully explain the 
pressure dependence of the Curie temperature, observed experimentally. 

Finally, we analyze the temperature dependence of the inverse uniform susceptibility 
in the mean-field approach and spin-fermion model (see explicit expressions in Appendix B). 
Our results rescaled with the number of orbitals are shown in the lower panel of 
Fig.~\ref{TC_SF} in comparison with the experimental data and the results of DMFT for 
the single-band model (Fig. \ref{Fig_7}). We observe that in the region $T>2000$~K the 
inverse susceptibility computed by DMFT becomes almost indistinguishable with the 
mean-field result. This shows correctness of account of screening of the interaction 
by choosing a reduced value of the Coulomb interaction in the mean-field approach 
(cf. Ref.~\cite{KataninYamase}) and confirms a view on the $T>T'_{\rm CW}$ state 
as on the weakly correlated state. Although at lower temperatures the susceptibility 
calculated by DMFT is suppressed by local correlations, these correlations, similarly 
to the results of the multi-band model in Fig. \ref{Fig_6}, are not sufficient to describe 
the experimental data. At the same time, the result of calculation of uniform magnetic 
susceptibility within the considered spin-fermion model shows further suppression of 
susceptibility by non-local correlations at moderate temperatures, but at the same time, 
its enhancement (in comparison to DMFT) in the low-temperature region. The resulting 
inverse susceptibility closely follows the experimental data. This shows again sufficiency 
of account of almost uniform classical fluctuations to achieve a good agreement with 
experiment.

\section{Conclusion}
\label{sec_conclusion}
In conclusion, by employing a combination of the DFT+DMFT method and analytical 
model-based techniques we investigated the electronic and magnetic properties of 
a prototypical itinerant ferromagnet ZrZn$_{2}$. Our DFT+DMFT results show that 
electronic correlations in ZrZn$_{2}$ are rather weak but not negligible as 
characterized by an effective mass renormalization of the Zr-$4d$ states $m^*/m\sim 1.1 - 1.3$. 
Most importantly, we demonstrate that the $4d$ $t_{2g}$ electronic states of Zr are 
partly localized by Coulomb interaction, yielding an orbital-selective formation of 
local magnetic moments, which are screened at low temperatures.

We also show that the effect of local Coulomb correlations is essential for determining 
the correct topology of some of the Fermi surface sheets. The shape of the Fermi surface 
computed by DFT+DMFT supports antiferromagnetic correlations, which compete with tendency 
to ferromagnetism and may explain a change of the temperature dependence of resistivity 
from $T^{5/3}$ to $T^{3/2}$ accompanying the destruction of ferromagnetic order. 

Extrapolation of the linear high-temperature part of the inverse uniform susceptibility 
$\chi^{-1}(T)$ of ZrZn$_2$ calculated by DFT+DMFT to $\chi^{-1}(T)=0$ yields an estimate 
of the characteristic temperature of the onset of strong ferromagnetic correlations 
$T_{\rm C}^{*}\sim 32$~K at ambient pressure and $T_{\rm C}^{*}\sim 4$~K at $p=2$~GPa 
in qualitative agreement with experimental behavior of the Curie temperature. 

We would like to emphasize, that the peaks of the density of states close to the 
Fermi level are rather common feature of many metallic compounds and they can 
significantly influence the physical properties~\cite{Katsnelson}. In general, 
these peaks on one hand enhance the tendency to ferromagnetic ordering, but on the 
other hand, they promote formation of the local moment and the corresponding Kondo 
screening at low temperatures. The offset of the peak of density of states from the 
Fermi level increases the corresponding Kondo temperature \cite{Skornyakov_PRB_2012,a-iron,e-iron,LaFeAsO,OurGamma}, and, therefore, serves 
as another source of suppression of ferromagnetism, apart from the suppression of 
the density of states. 

Accordingly, at low temperatures the shape of $\chi^{-1}(T)$, computed by DFT+DMFT, 
shows a significant deviation from the linear behavior (an upturn). In view of the 
obtained partial localization of the Zr $4d$ $t_{2g}$ states this upturn may be 
explained by a competition of the ferromagnetic ordering with the Kondo effect, which is similar to that considered previously for Kondo lattices ~\cite{CompFMKondo1,Ohkawa,Si,CompFMKondo2,Kubo,Hoshino}. This competition 
was also considered a possible source of weak ferromagnetism ~\cite{IrkhinKondo1}, and it 
is expected to be quite sensitive to spatial correlations.

Importance of the Kondo physics for  of ZrZn$_2$ is supported by the observation 
that the shape of $\chi(T)$ obtained by DFT+DMFT is similar to that of FeAl where 
ferromagnetism is destroyed by the Kondo screening~\cite{FeAl}. While the peak of 
the density of states of ZrZn$_2$ is closer to the Fermi level, than that for FeAl, 
DFT+DMFT for ZrZn$_2$ also yields a paramagnetic ground state, although with a 
lower temperature of the susceptibility upturn. Since in case of a single impurity 
the Kondo temperature is typically larger than the Kondo coherence temperature 
for the lattice, the paramagnetic state in DMFT for ZrZn$_2$ may be a consequence 
of an overestimate of the Kondo effect by this approach. A possible reason of this 
overestimate is that suppression of the Kondo effect by non-local magnetic correlations 
is not accounted by DMFT approach. While the upturn of calculated inverse susceptibility 
in ZrZn$_2$ can be explained in terms of the Kondo screening, which is overestimated 
by DMFT, the effect of the approximations employed within the CT-QMC method, namely 
Ising-type Hund exchange and the neglect of the effect of spin-orbit coupling, 
requires further consideration. 

We note also that approaching (and even underestimating) the experimental Curie 
temperature by DFT+DMFT  
is obtained for such a simple magnet as nickel. 
Specifically, in this case DMFT approach with Ising (or SU(2)) Hund exchange yields $T_{\rm C}=700$~K (or $600$~K) \cite{Lichtenstein2001,Hausoel} vs. the experimental $T_{\rm C}=630$~K. 
The effective spin of Ni $p=0.45$ as determined via Eq. (\ref{Effspin}) using the 
experimental magnetic moment $\mu=1.6\mu_{\rm B}$ appears to be slightly less than 
$1/2$. This along with the proximity of the Kondo temperature $T_{\rm K}\approx 800$~K, 
extracted from the local susceptibility of Ref. \cite{Hausoel}, to the Curie 
temperature, also conforms to the presence of a weak Kondo screening in this magnet 
and shows a possibility of underestimation of the Curie temperature within DFT+DMFT.

To study details of the paramagnet-ferromagnet transition in ZrZn$_2$, we have 
constructed an effective single-band model by using the Zr $4d$ contribution to 
the total density of states. We have shown that for an appropriate choice of the 
Coulomb repulsion DMFT solution of this model yields results, close to those  
of the realistic multi-band model.

Our results for the effective single-band model obtained within the static mean-field 
approximation show a second-order paramagnet-ferromagnet phase transition with 
decreasing temperature at small pressures. Upon increasing pressure the first-order 
quantum phase transition into the paramagnetic phase occurs in agreement with 
experimental data. Yet, in contrast to DMFT the static mean-field approximation 
overestimates the transition temperature by almost an order of magnitude. 

To reconcile the results of static and dynamic mean-field approaches to the 
single-band model, we have also considered a spin-fermion approach to the 
single-band model. Similarly to DMFT, this approach yields reasonable values 
of transition temperatures and, at the same time, it preserves the second-order 
phase transition with changing temperature at low pressures and the first-order 
quantum phase transition into paramagnetic phase at elevated pressures. Within 
this model computed pressure dependence of the Curie temperature and the temperature 
dependence of susceptibility at ambient pressure are in good quantitative agreement 
with experimental data. 

The considered spin-fermion model relies on consideration of ferromagnetic correlations, 
putting aside other types of correlations and possible Kondo effect (which should be 
undoubtedly studied in the future). On one hand this approach allows us to obtain 
physically valuable information about ferromagnetic properties of ZrZn$_2$. 
Therefore, it is perspective to study magnetic properties of other weak itinerant 
magnets. On the other hand, this calls for further development of  methods 
describing both, local and non-local correlations in weak magnets. These approaches 
can be based on diagrammatic non-local extensions of DMFT \cite{OurRevModPhys}.
Correct treatment of Kondo effect in ZrZn$_2$ may also require account of the non-local 
correlations within the Kondo lattice or periodic Anderson model. Despite these models 
were intensively applied to describe magnetism of heavy-fermion systems (see, e.g., 
Ref.~\cite{KondoLattice}), their possible relevance for weak magnets with peaks of 
the density of states (cf. Ref. \cite{IrkhinKondo1}) has to be further explored. 

\section{ACKNOWLEDGMENTS}
The DFT+DMFT calculations were supported by the Russian Science Foundation 
(Project 19-12-00012). The dynamical mean-field calculations of the one-band 
model were performed within the state assignment of Minobrnauki of Russia 
(theme Electron No. AAAA-A18-118020190098-5). The study of the one-band model 
within static mean-field theory and spin-fermion approach was supported by 
RFBR grant 17-02-00942a. A.~A.~K. acknowledges V. Yu. Irkhin for discussions on 
the properties of weak magnets.

\appendix
\section{Uniform susceptibility at temperatures higher the width of the peak}
We consider the uniform static susceptibility of the multi-band model in the 
form of a single bubble, Eq. (\ref{Eq:Bubble}) at ${\bf q}=0$,
\begin{eqnarray}
\chi_{0}=-{2\mu_B^2 T}\sum_{{\bf k},i\nu_n,m,m'}G_{{\bf k},mm'}(i\nu_n)G_{{\bf k},m'm}(i\nu_n),\label{bubble}
\end{eqnarray} 
where $G_{{\bf k},mm'}=\langle c^+_{k,m,\sigma}(0) c_{k,m',\sigma}(\tau) \rangle_{i \nu_n}$
is the electronic Green function, $c^+_{k,m,\sigma}$, $c_{k,m',\sigma}$ are the 
electron creation and destruction operators with momentum ${\bf k}$, orbital $m$ 
and spin projection $\sigma$. Using the spectral representation for the Green's 
functions we find
\begin{eqnarray}
\chi_{0}=&-&{2\mu_B^2 T}\sum_{i\nu_n,m,m'}\frac{A_{{\bf k},mm'}(\nu) A_{{\bf k},m'm}(\nu')}{(i\nu_n-\nu)(i\nu_n-\nu')}\\
&=&{2\mu_B^2}\sum_{{\bf k},m,m'}\int d\nu \int d\nu' {A_{{\bf k},mm'}(\nu) A_{{\bf k},m'm}(\nu')}\notag\\
&\times&\frac{f(\nu)-f(\nu')}{\nu'-\nu}\\
&=&{2\mu_B^2}\sum_{{\bf k},m,m'}\int d\nu \int d\nu' {A_{{\bf k},mm'}(\nu) A_{{\bf k},m'm}(\nu')}\notag\\
&\times&\left[1-f(\nu)\right]f(\nu')\frac{\exp[(\nu'-\nu)/T]-1}{\nu'-\nu}\label{chi0Eq}
\end{eqnarray} 
where $A_{{\bf k},mm'}(\nu)=(-1/\pi){\rm Im}G_{{\bf k},mm'}(\nu)$ is the spectral 
weight of the Green's function, $f(\nu)$ is the Fermi function. The $1/T$ contribution 
to susceptibility originates from the integration region $|\nu-\nu'|\ll T$. Together 
with $\nu\gtrsim -T$ and $\nu'\lesssim T$ restricted by the Fermi functions, this 
implies $|\nu|,|\nu'|\lesssim T$ in that region. 

Let us consider the contribution to the susceptibility from some relevant energy 
range $\nu,\nu'\in[\epsilon_{\rm min},\epsilon_{\rm max}]$, which contains peaks of 
the density of states near the Fermi level. In our case this refers to Zr 4d states. 
The reason is that such contributions, resummed within the random-phase approximation, 
provide the Curie-Weiss law up to relatively low temperatures 
$T\gtrsim \max(|\epsilon_{\rm min}|,\epsilon_{\rm max})$. Indeed, at 
$T\gg \max(|\epsilon_{\rm min}|,\epsilon_{\rm max})$ we have  
$|\nu-\nu'|<\epsilon_{\rm max}-\epsilon_{\rm min}\ll T$, such that the abovementioned 
$1/T$ behavior of the bubble takes place. We obtain for the considered contribution
\begin{eqnarray}
\chi_{0}&=&\frac{2\mu_B^2}{T}\sum_{{\bf k},m,m'}\int d\nu \int d\nu' {A_{{\bf k},mm'}(\nu) A_{{\bf k},m'm}(\nu')}\notag\\
&\times&\left[1-f(\nu)\right]f(\nu')\\
&=&\frac{2\mu_B^2}{T}\sum_{{\bf k},m,m'} \langle c_{{\bf k},m}c^+_{{\bf k},m'}\rangle \langle c^+_{{\bf k},m'}c_{{\bf k},m}\rangle\label{Eqchi1}
\end{eqnarray}
In the considered temperature range the averages in the last line weakly depend 
on $\bf k$ and their local value can be taken, which yields
\begin{equation}
    \chi_{0}=\frac{4\mu_B^2}{3T}\langle S^2\rangle=\frac{\mu^2}{3T}\label{Eqchi2}
\end{equation}
where $\langle S^2\rangle$ is the average square of spin on the relevant orbitals 
($4d$ Zr in our case). For ZrZn$_2$ we obtain from DFT+DMFT calculations of the 
multiorbital model $\langle S^2\rangle=1.43$. This yields  corresponding magnetic 
moment $\mu=2.39\mu_B$, which agrees well with the moment extracred from inverse 
susceptibility in the temperature range $T>T'_{\rm CW}$ (see Sect. II B 4). Note that 
obtained in DFT+DMFT intraorbital contributions to this value,
$\mu=((3/2)\sum_m n_m(2-n_m))^{1/2}\mu_B=2.16\mu_B$, so that the interorbital 
correlations are relatively weak.

The results (\ref{Eqchi1}) and (\ref{Eqchi2}) can be extended to the uniform 
susceptibility in DMFT, which can be represented in the form \cite{dftdmft_method1,OurRevModPhys}
\begin{equation}
{\chi}_{\mathbf{q}=0}^{\rm DMFT}=2 \mu_B^2 \sum_{\substack
{\nu,~m,~m',\\\nu',m'',m'''}
} \left[
\delta_{\nu\nu'}\hat{\chi}^{-1}_{0,\nu}- \hat{\Gamma} \right]
_{\substack
{\nu~m~m',\\\nu'm''m'''}
}
^{-1} \label{EqFq}
\end{equation}
where $\hat{\chi}_0$ is the matrix in orbital indexes of the bubbles of Green's 
functions, similar to the Eq. (\ref{bubble}),
\begin{eqnarray}
\hat{\chi}_{0,\nu}^{mm',m''m'''}=-T \sum_{\bf k} G_{{\bf k},mm''}(i\nu)G_{{\bf k},m'''m'}(i\nu),\notag
\end{eqnarray} 
and $\hat{\Gamma}$ is the irreducible vertex (which is also considered as a 
matrix in frequency- and orbital space). Assuming that the vertex $\Gamma$ is 
weakly frequency- and orbital dependent, using spectral representation for 
Green's functions, which is similar to that in Eq. (\ref{chi0Eq}) and considering 
the contribution of the energy range $|\nu|,|\nu'|\lesssim T$, we arrive at 
the random-phase approximation (RPA)-like results for the susceptibility, which fulfills the Curie-Weiss law, 
with the Curie constant given approximately by Eq. (\ref{Eqchi2}). The main effect 
of the vertex $\Gamma$ in this case is in the shift of the inverse susceptibility 
with respect to the Curie law (\ref{Eqchi2}), i.e. introducing a finite Weiss 
temperature. 

\section{Derivation of the equations of the spin-fermion model}
The action (\ref{action}) can be written as quadratic form of fermionic operators as follows:
\begin{equation}
\mathcal{S}_{{\rm eff}}=\sum_{k,\sigma,\sigma^{'}}c_{k,\sigma}^{\dagger}M_{\sigma,\sigma^{'}}(k)c_{k,\sigma^{'}}+D\tilde{\textbf{S}}^{2}+ E_0,
\end{equation}
where $E_0=U(\langle s_z \rangle^2-\langle n \rangle^2/4)$,
\begin{widetext}
\begin{equation}
\textbf{M}=
\begin{pmatrix}
-i{ \nu_{n}}+\tilde{\epsilon}_{\textbf{k}}-{H_{\rm MF}}/{2}+U\tilde{S}_z& U\left(\tilde{S}_x-i\tilde{S}_y\right)\\
U\left(\tilde{S}_x+i\tilde{S}_y\right)&-i{ \nu_{n}}+\tilde{\epsilon}_{\textbf{k}}+{H_{\rm MF}}/{2}-U\tilde{S}_z
\end{pmatrix},
\label{M}
\end{equation}
\end{widetext}
$H_{\rm MF}=H+2U\langle s_{z}\rangle$, we have added the magnetic field 
$H$ for completeness, and
$$ 
\tilde{\epsilon}_{\textbf{k}}=
\epsilon_{\textbf{k}}-\mu+\dfrac{U\langle n \rangle}{2}.
$$
After integrating out the fermions in Eq.~(\ref{PF}), we get the partition 
function $Z$ in the form
\begin{multline}
Z=
e^{-{\beta}E_0}\int{d^{3}\tilde{S}\exp{\left(-{\beta}D\tilde{\textbf{S}}^{2}+\sum_{\textbf{k},n}{\ln{\left({\beta^{2}}\det{\textbf{M}}\right)}}\right)}}.
\label{P2}
\end{multline}
Using Eq.~(\ref{M}), we obtain
\begin{equation}
\det \textbf{M}=\left(-i{ \nu_{n}}+\tilde{\epsilon}_{\textbf{k}}\right)^{2}-\gamma^{2},
\end{equation}
where 
\begin{equation}
\gamma^{2}=
H_{\rm MF}^2/4-
U 
H_{\rm MF}
\tilde{S}_z+U^{2}\tilde{S}^{2}.
\end{equation}
The Matsubara sum in Eq.~(\ref{P2}) can be carried out exactly, yielding
\begin{eqnarray}
\sum_{\textbf{k},n}{\ln{\left({\beta^{2}}\det{\textbf{M}}\right)}}&=&\sum_{\textbf{k},n,\sigma}{\ln{\left({\beta}\left(-i{ \nu_{n}}+\tilde{\epsilon}_{\textbf{k}}+\sigma\gamma\right)\right)}}\notag\\&=&-\sum_{\textbf{k},\sigma}{\ln{\left(1-f(\tilde{\epsilon}_{\textbf{k}}+\sigma\gamma)\right)}}.
\end{eqnarray}
Substituting this result in Eq.~(\ref{P2}), the partition function finally becomes
\begin{eqnarray}
Z&=&e^{-{\beta}E_0}\int{d^{3}}\tilde{S}\exp{\biggl(-{\beta}D\tilde{\textbf{S}}^{2}-R(\tilde{\bf S})\biggr)},
\label{Zf}
\end{eqnarray}
where 
$R(\tilde{\bf S})=\sum_{\sigma}{\int{d\epsilon\rho\left(\epsilon\right)\ln{\left(1-f(\tilde{\epsilon}+\sigma\gamma)\right)}}}$. 
Therefore, in the presented  spin-fermion model, the average total occupation 
$\langle n \rangle$ is given by
\begin{multline}
\langle n \rangle =-\dfrac{\partial \Omega}{\partial \mu}
=\dfrac{{e^{-{\beta}E_0}}}{Z}\int{d^{3}}\tilde{S}\exp{\biggl(-{\beta}D\tilde{\bf S}^{2}}-R(\tilde{\bf S})\biggr)\\
\times\left[\sum_{\sigma}{\int{f\left(\tilde{\epsilon}+\sigma\gamma\right)\rho(\epsilon)d\epsilon}}\right]
\label{Nt}
\end{multline}
and the average magnetization $\langle s_{z} \rangle$ is
\begin{multline}
\langle s_{z} \rangle=-\dfrac{\partial \Omega}{\partial H}=-\dfrac{{e^{-{\beta}E_0}}}{Z}\int{d^{3}}\tilde{S} 
\exp{\biggl(- {\beta}D\tilde{\textbf{S}}^{2}-\biggr.}\biggl.R(\tilde{\bf S})\biggr)
\\\times\left[\sum_{\sigma}{\sigma\int{\dfrac{\dfrac{H_{\rm MF}}{2}
-U\tilde{S}_z}{2\gamma}f\left(\tilde{\epsilon}+\sigma\gamma\right)\rho(\epsilon)d\epsilon}}\right]
\label{Mt}
\end{multline}
where $\Omega=-{\beta^{-1}}\ln Z$ is the thermodynamic potential.

The expansion of logarithmic contribution $R(\tilde{\bf S})$ in the exponents of Eqs.~ (\ref{Zf})-(\ref{Mt}) 
in powers of $\gamma$ contains only even powers and corresponds to weak 
magnetization of the spin-fluctuation field (the first order in $\tilde{S}$ term), 
the renormalization of the value of $D$ (for the second-order term) and 
multi-paramagnon interactions (higher-order terms). We neglect the weak 
magnetic field acting on $\tilde{S}$ and multi-paramagnon interactions 
effects and assume that the value of $D$ is already renormalized. These 
approximations imply replacement of $\gamma$ in these logarithmic contributions 
by its $\tilde{S}=0$ value. At fixed 
$\langle n \rangle=2\int_{-\infty}^{\epsilon_{F}}{\rho(\epsilon)d\epsilon}$, $H=0$, $T$, $D$, 
the equations (\ref{Nt}) and (\ref{Mt}) reduce therefore to the equations (\ref{SF_EQn}) and 
(\ref{SF_EQm}) of the main text. 

To calculate the susceptibility we differentiate Eq. (\ref{Mt}), neglecting 
the logarithmic contribution in the exponent, over $H$. Using
\begin{equation}
\frac{dH_{MF}}{dH}=1+2U\chi_s,    
\end{equation}
where $\chi_s=d\langle s_z\rangle/dH$, we find at $H\rightarrow 0$
\begin{eqnarray}
&&\frac{d\gamma}{dH}=\frac{U}{2\gamma}
\left(\langle s_{z}\rangle-\tilde{S}_{z}\right)\left(1+2U\chi_s\right),\\
&&\frac{d}{dH}\left(\dfrac{{H_{\rm MF}}/{2}
-U\tilde{S}_z}{2\gamma}\right)\notag=U^{2}\dfrac{\tilde{S}_{x}^{2}+\tilde{S}_{y}^{2}}{4\gamma^{3}}\left(1+2U\chi_s\right)
\end{eqnarray}
In the following we assume paramagnetic phase $\langle s_z\rangle=0$. 
In this case the derivative $d\mu/dH$ does not contribute to the susceptibility 
$\chi$ since the corresponding term is odd in $\tilde{S}_z$. Collecting all 
the terms together we obtain RPA-like result
\begin{equation}
\chi=4\mu_B^2\chi_s=2\mu_B^2\dfrac{\chi_{0}^{\rm SF}}{1-U\chi_{0}^{\rm SF}},\label{RPA}
\end{equation}
where
\begin{equation}
\chi_{0}^{\rm SF}=(2Z_{0})^{-1}\sum_{\sigma}\left(F^{\sigma}-\sigma D^{\sigma}\right),
\label{chi_0}
\end{equation}
we have introduced
\begin{equation}
F^{\sigma}
=\dfrac{4\pi}{3}\int_{0}^{\infty}{d\tilde{S}}\tilde{S}^{2}B^{\sigma}(\tilde{S})\exp{(- {\beta}D\tilde{\textbf{S}}^{2})}
,  
\label{Fsigma}
\end{equation}
\begin{equation}
D^{\sigma}
=\dfrac{8\pi}{3U}\int_{0}^{\infty}{d\tilde{S}}\tilde{S}A^{\sigma}(\tilde{S})\exp{(- {\beta}D\tilde{\textbf{S}}^{2})}
,  
\label{Dsigma}
\end{equation}
and 
\begin{eqnarray}
A^{\sigma}(\tilde{S})&=&\int{f\left(\tilde{\epsilon}+\sigma\gamma\right)\rho(\epsilon)d\epsilon},\\
B^{\sigma}(\tilde{S})&=&-\int{ \dfrac{\partial f\left(\tilde{\epsilon}+\sigma\gamma\right)}{\partial \epsilon}\rho(\epsilon)d\epsilon},
\end{eqnarray}
\begin{equation}
Z_{0}=\int{d^{3}}\tilde{S}\exp{(- {\beta}D\tilde{\textbf{S}}^{2})}.
\end{equation}
Integrating by parts, equation~(\ref{Dsigma}) can be written as 
\begin{eqnarray}
D^{\sigma}&=&\dfrac{4\pi}{3\beta D U}\biggl[A^{\sigma}(0)\biggr. \notag\\
&-&\left.\sigma U\int_{0}^{\infty}{d\tilde{S}} B^{\sigma}(\tilde{S})\exp{(- {\beta}D\tilde{\textbf{S}}^{2})}\right]
\label{Dsigma_2}
\end{eqnarray}
Substituting Eqs.~(\ref{Dsigma_2}) into Eq.~(\ref{chi_0}), we obtain
\begin{eqnarray}
\chi_{0}^{\rm SF}&=&\dfrac{2\pi}{3Z_{0}}\int_{0}^{\infty}{d\tilde{S}}\left(\tilde{S}^{2}+\dfrac{1}{\beta D}\right)\notag\\
&\times&\sum_{\sigma}B^{\sigma}(\tilde{S})\exp{(- {\beta}D\tilde{\textbf{S}}^{2})}
\end{eqnarray}
In the absence of spin fluctuations ($D\rightarrow \infty$), i.e. in mean-field 
approach, we find RPA result of Eq. (\ref{RPA}) with 
$\chi_{0}^{\rm SF}\rightarrow -\int \rho(\epsilon)f'(\tilde{\epsilon})d\epsilon$, 
which is the bare spin susceptibility. The resulting temperature dependence of 
the inverse susceptibility in the mean-field and spin-fermion model is discussed 
in Sect. IIIC.

\end{document}